\documentclass[aps,prl,twocolumn,showpacs,superscriptaddress]{revtex4-1}
\usepackage{bm}
\usepackage{mathrsfs}
\usepackage{amsmath}
\usepackage{amssymb}
\usepackage{graphicx}
\usepackage{amsfonts}
\usepackage{amsthm}
\usepackage{color}
\usepackage{dcolumn}
\usepackage{txfonts}
\usepackage{epsfig}

\begin{document}
\title{Relations between Dissipated Work and R\'{e}nyi Divergences}
\author{Bo-Bo Wei}
\affiliation{Institut f\"{u}r Theoretische Physik, Albert-Einstein-Allee 11, Universit\"{a}t Ulm, 89069 Ulm, Germany}
\author{M. B. Plenio}
\affiliation{Institut f\"{u}r Theoretische Physik, Albert-Einstein-Allee 11, Universit\"{a}t Ulm, 89069 Ulm, Germany}

\begin{abstract}
In this paper, we establish a general relation which directly links the dissipated work done on a system driven arbitrarily far from equilibrium, a fundamental quantity in thermodynamics, and the R\'{e}nyi divergences, a fundamental concept in information theory. Specifically, we find that the generating function of the dissipated work under an arbitrary time-dependent driven process is related to the R\'{e}nyi divergences between a non-equilibrium state in the driven process and a non-equilibrium state in its time reversed process. This relation is a consequence of time reversal symmetry in driven process and is universally applicable to both finite classical system and finite quantum system, arbitrarily far from equilibrium.
\end{abstract}
\maketitle

The pioneering works by Clausius and Kelvin have established that the
average mechanical work needed to move a system in
contact with a heat bath at temperature $T$, from one equilibrium
state $A$ into another equilibrium state $B$, is at least
equal to the free energy difference between these states:
$\langle W\rangle\geq F_B-F_A$, where the equality holds only for a quasi-static
process. In a remarkable development Jarzynski \cite{Jarzynski1997} discovered that for a classical system initialized in an equilibrium state the work done under a non-equilibrium change of control parameters is related to the equilibrium free energy difference between the initial and the final equilibrium states for the control parameters via
\begin{eqnarray}\label{JE}
\langle e^{-\beta W}\rangle=\frac{Z(\beta,\lambda_f)}{Z(\beta,\lambda_i)}=e^{-\beta [F(\beta,\lambda_f)-F(\beta,\lambda_i)]}.
\end{eqnarray}
Here $\beta\equiv 1/T$ is the inverse temperature $T$ of the initial equilibrium state of the system and we take the Boltzmann constant $k_B\equiv1$, $W$ is the work done on the system due to a driving protocol under which the control parameter changes from $\lambda_i$ to $\lambda_f$, $F$ is the Helmholtz free energy of the system and is defined by $F=-k_BT\ln Z$ with $Z$ being the equilibrium partition function and the angular bracket on the left of Equation \eqref{JE} denotes an ensemble average over realizations of the process. The Jarzynski equality connects equilibrium thermodynamic quantity, the free energy difference, to a non-equilibrium quantity, the work done in a processes that may be carried out arbitrarily far from equilibrium. It implies that we can determine the equilibrium free energy difference of a system by repeatedly performing work at any rate. Jarzynski equality and Crooks relation \cite{Crooks1999} from which it can be derived have been verified experimentally in various physical systems \cite{PNAS2001,Science2005,Nature2005,EPL2005,PT2005,PRL2006,PRL2007,PRL2012} and were also proved to hold for finite quantum mechanical systems \cite{arXiv2000a,arXiv2000b,Talker2007,Kim2015} provided that work in a quantum system is defined by two projective measurements \cite{Talker2007}. The discovery of the Jarzynski equality has led to a very active field concerned with fluctuation relations in non-equlibrium thermodynamics \cite{RMP2009,Jar2011,RMP2011}.

The excess work $W-\Delta F$ that arises in irreversible processes is often referred to as the dissipated work, $W_{\text{diss}}=W-\Delta F$. In terms of the dissipated work, Jarzynski equality can be written as \cite{Jarzynski1997},
\begin{eqnarray}\label{JE1}
\langle e^{-\beta W_{\text{diss}}}\rangle=1.
\end{eqnarray}
It is an identity that provides constraints on the dissipated work in an arbitrarily driven process. Contrary to the reversible work, which only depends on the initial and final equilibrium states,
the dissipated work depends on how the specific driven protocol is performed.  Usually the driven protocol is realized by changing the control parameters in the Hamiltonian between initial and final values, which can in
principle bring the system arbitrarily far out of equilibrium.
Surprisingly, there exists a neat and exact
microscopic fluctuation relation for the dissipated work. The central
result of this paper is the following relation:
\begin{eqnarray}\label{central}
\langle \Big(e^{-\beta W_{\text{diss}}}\Big)^z\rangle =e^{(z-1)S_{z}[\Theta\rho_R(\tau-t)\Theta^{-1}||\rho_F(t)]},
\end{eqnarray}
where $z$ is a finite real number, $W_{\text{diss}}$ is the dissipated work done on the system due to a driving protocol under which the control parameter changes from $\lambda_i$ to $\lambda_f$ in time duration $\tau$, the angular bracket on the left hand side denotes an ensemble average over the realizations of the driven process and $S_{z}[\Theta\rho_R(\tau-t)\Theta^{-1}||\rho_F(t)]$ is the order-$z$ R\'{e}nyi divergence between $\Theta\rho_R(\tau-t)\Theta^{-1}$ and $\rho_F(t)$ with $\Theta$ being the time reversal operation. For classical system, the order-$z$ R\'{e}nyi divergence for distributions $\rho_1(X)$ and $\rho_2(X)$ is defined as \cite{Renyi1961,Erven2014}  $S_{z}[\rho_1||\rho_2]\equiv\frac{1}{z-1}\ln[\int dX\rho_1^{z}(X)\rho_2^{1-z}(X)]]$. $\rho_F(t)$ is the phase space
density in the forward driven process measured at an arbitrary intermediate time $t$ which was initialized in the canonical equilibrium state at inverse temperature $\beta$ and control parameter $\lambda_i$ and $\rho_R(\tau-t)$ is the phase space
density in the time reversed driven process measured at intermediate time $\tau-t$ which was initialized in the canonical equilibrium state at inverse temperature $\beta$ and control parameter $\lambda_f$. For quantum system, the order-$z$ R\'{e}nyi divergence for quantum states $\rho_1$ and $\rho_2$ is defined as \cite{Beigi2013,Lennert2013} $S_{z}[\rho_1||\rho_2]\equiv\frac{1}{z-1}\ln[\text{Tr}[\rho_1^{z}\rho_2^{1-z}]]$. $\rho_F(t)$ is the density matrix of the system measured at an arbitrary intermediate time $t$ in the forward driven process which was initialized in the canonical equilibrium state at inverse temperature $\beta$ and force parameter $\lambda_i$ and $\rho_R(\tau-t)$ is the density matrix at time $\tau-t$ in the time reversed driven process which was initialized in the canonical equilibrium state at inverse temperature $\beta$ and force parameter $\lambda_f$.

Since the concept of work in classical system and quantum system are subtly different \cite{Talker2007,RMP2011}, in the following we shall separately discuss derivation of Equation \eqref{central} in non-equilibrium classical thermodynamics and in non-equilibrium quantum thermodynamics.

\emph{Non-equilibrium classical thermodynamics}-
We consider a finite classical system with Hamiltonian $H[X;\lambda]$, where $X=[q_1,p_1;q_2,p_2;\cdots;q_N,p_N]$ denotes collectively the coordinates and momenta of all the $N$ particles in the system, $\lambda$ is a parameter controlled by an external agent. For a classical system with time-dependent Hamiltonian, the microscopic reversibility \cite{Stra1994} is illustrated in Figure 1.

\begin{figure}
\begin{center}
\includegraphics[scale=0.25]{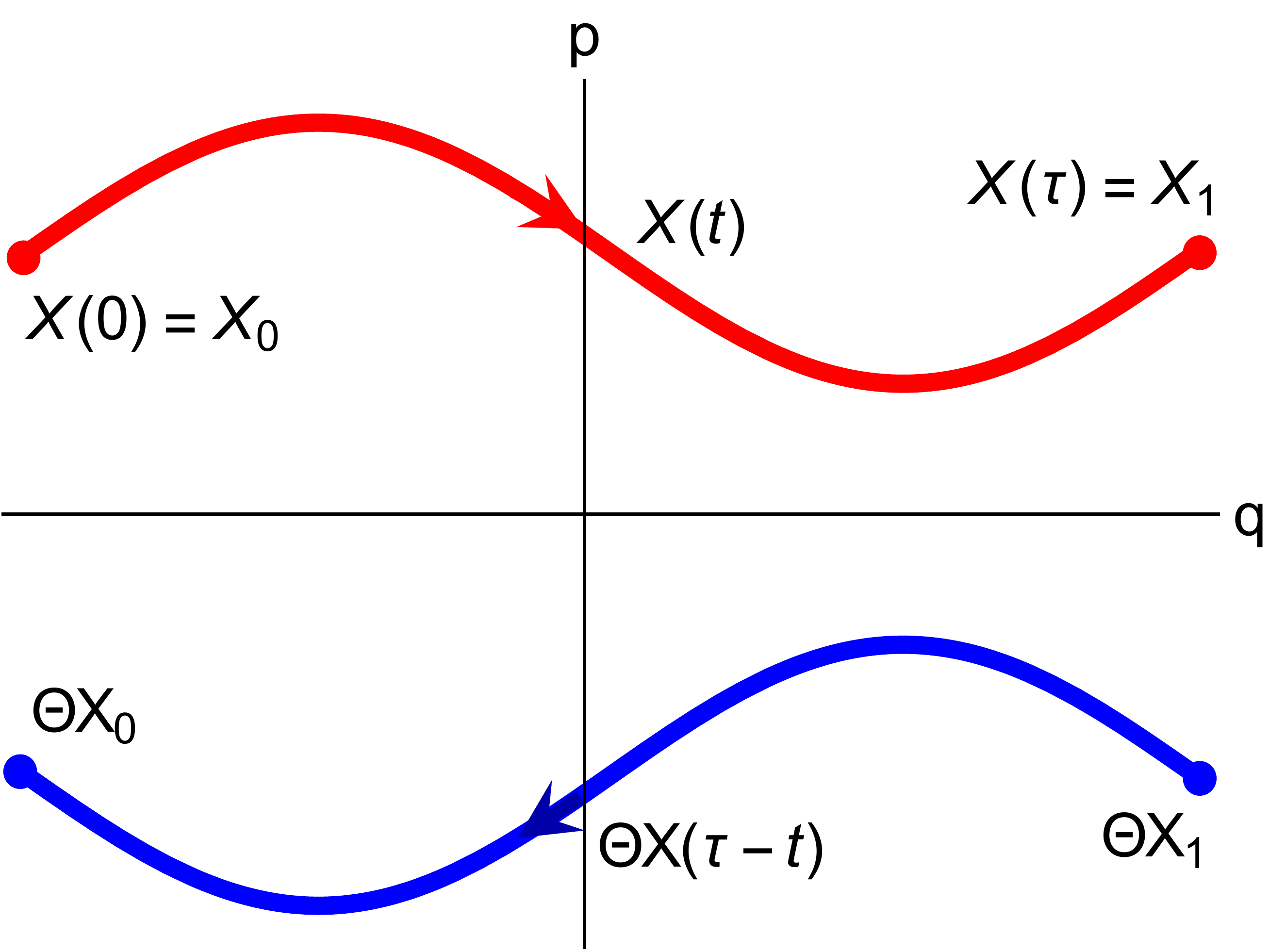}
\end{center}
\caption{(color online). Schematic illustration of the time reversal symmetry for classical system with time-dependent Hamiltonian. The upper red line with arrow towards right is a trajectory in the forward driven process, it starts at $X_0$ and is driven by $H[X,\lambda(t')], t'\in[0,\tau]$ and arrives in $X(t)$ at time $t$ and finally ends in $X_1$ at time $\tau$. The lower blue line with arrow towards left is its time reversed trajectory, which starts at $\Theta X_1$ (time reversed state of $X_1$) and is driven by $H_R[X;t]=\Theta H[X;\lambda(\tau-t)]\Theta^{-1}, t'\in[0,\tau]$ and arrives in $\Theta X(\tau-t)$ after time $t$ and finally returns to $\Theta X_0$ at time $\tau$. }
\label{fig:epsart1}
\end{figure}

We first introduce the \emph{forward process} for a classical system under time-dependent driving. We assume that the classical system $H[X;\lambda]$ is initialized in a canonical equilibrium state at inverse temperature $\beta=1/T$ at the value $\lambda_i$ of the control parameter, which is described by the Boltzmann-Gibbs distribution in phase space, $\rho_F[X;0]=e^{-\beta H[X;\lambda_i]}/Z(\beta,\lambda_i)$ with $Z(\beta,\lambda_i)=\int dX_0 e^{-\beta H[X_0;\lambda_i]}$ being the initial partition function. Then the classical system is isolated and driven by an external agent, which varies the control parameter $\lambda$ from an initial value $\lambda_i$ to a final value $\lambda_f$ in a time duration $\tau$ according to a specified protocol $\lambda(t), t\in[0,\tau]$. Then the phase space density evolves in time under the Liouville equation \cite{Reichl1987},
\begin{eqnarray}
\partial_t\rho_F[X;t]=\{H[X;\lambda(t)],\rho_F[X;t]\},
\end{eqnarray}
where $\{A,B\}\equiv\sum_{j=1}^N[\frac{\partial A}{\partial q_j}\frac{\partial B}{\partial p_j}-\frac{\partial A}{\partial p_j}\frac{\partial B}{\partial q_j}]$ is the Poisson bracket in the Hamilton mechanics. Of course usually under time dependent driven, $\rho_F[X;t]\neq e^{-\beta H[X;\lambda(t)]}/Z[\beta,\lambda(t)]$. However the Liouville theorem states that the phase space distribution is invariant along any trajectory of the system \cite{Reichl1987}, thus one has, for $\forall t\in[0,\tau]$,
\begin{eqnarray}\label{classicalforward}
\rho_F[X;t]=\rho_F[X_0;0],
\end{eqnarray}
where $X$ is the resulting phase space point at time $t$ under the dynamics of forward Hamiltonian $H[X;\lambda(t)]$ if it was initially at $X_0$ at $t=0$ [See the upper red line in Figure 1]. According to first law of
thermodynamics, the work done associated with the trajectory in the forward process [The upper red in Figure 1] only depends on the initial state if the force protocol is fixed and we have
\begin{eqnarray}\label{work}
W[X_0]=H[X_1;\lambda_f]-H[X_0;\lambda_i].
\end{eqnarray}

Now we consider the \emph{reversed process}. In the reversed process, the classical system is initialized in a canonical equilibrium state at inverse temperature $\beta$ at the value $\lambda_f$ of the control parameter, $\rho_R(X;0)=e^{-\beta H_R[X;0]}/Z(\beta,\lambda_f)=\Theta e^{-\beta H[X;\lambda_f]}\Theta^{-1}/Z(\beta,\lambda_f)$ with $Z(\beta,\lambda_f)=\int dX e^{-\beta H[X;\lambda_f]}$ being the initial partition function in the reversed process. Then the classical system is completely isolated and is driven by the reversed Hamiltonian $H_R[X;t]=\Theta H[X;\lambda(\tau-t)]\Theta^{-1}, t\in[0,\tau]$ for a time duration $\tau$. The dynamics of the phase space density for the reversed process at time $t$ is governed by Liouville equation \cite{Reichl1987}
\begin{eqnarray}
\partial_t\rho_R[X;t]=\{H_R[X;t],\rho_R[X;t]\}.
\end{eqnarray}
Of course usually under time dependent driven $\rho_R[X;t]\neq e^{-\beta H[\Theta X;\lambda(\tau-t)]}/Z[\beta,\lambda(\tau-t)]$. However the Liouville theorem states that, for $\forall t\in[0,\tau]$,
\begin{eqnarray}\label{classicalreversed}
\rho_R[\Theta X(\tau-t);t]=\rho_R[\Theta X_1;0].
\end{eqnarray}
Here $\Theta X(\tau-t)$ is the resulting phase space points at time $t$ under the dynamics of Hamiltonian in the reversed process $H_R[X;t]$ if it was at $\Theta X_1$ at $t=0$ [See the lower blue line in Figure 1].

Combing Equation Equation \eqref{classicalforward}, \eqref{work} and \eqref{classicalreversed}, we obtain
\begin{eqnarray}\label{crooks1}
e^{-\beta (W-\Delta F)}=\frac{\rho_R[\Theta X(t);\tau-t]}{\rho_F[X(t);t]}=\frac{\Theta\rho_R[X(t);\tau-t]\Theta^{-1}}{\rho_F[X(t);t]}.
\end{eqnarray}
Here $\Delta F\equiv F[\beta,\lambda_f]-F[\beta,\lambda_i]$, $t\in[0,\tau]$ is arbitrary time points. Note that on the right hand side of Equation \eqref{crooks1}, the phase space densities are observed at the same phase space point $X(t)$. This is a special case of the generalized Crooks relations for classical systems \cite{Kawai2007}. It states that the work done associated with a trajectory in phase space is fully determined by the phase space density in the forwarded process at any intermediate time and the phase space density of its time reversed process at arbitrary intermediate time. It is a consequence of Liouville theorem in classical mechanics.

Making use of Equation \eqref{crooks1}, we have
\begin{eqnarray}
\langle \Big(e^{-\beta W}\Big)^z\rangle_F\
&=&\int dX_0 \rho_F(X_0;0)e^{-\beta zW[X_0]},\label{c1}\\
&=&e^{-\beta z\Delta F}\int dX \rho_F(X;t)\Bigg(\frac{\rho_R(\Theta X;\tau-t)}{\rho_F(X;t)}\Bigg)^z,\label{c2}\\
&=&e^{-\beta z\Delta F}\int dX\rho_F(X;t)^{1-z}\rho_R(\Theta X;\tau-t)^z,\\
&=&e^{-\beta z\Delta F}e^{(z-1)S_z[\rho_R(\Theta X;\tau-t)||\rho_F(X;t)]}\label{c4},
\end{eqnarray}
where $z$ is a finite real number and $S_z[\rho_1||\rho_2]\equiv\frac{1}{z-1}\ln[\int dX\rho_1(X)^z\rho_2(X)^{1-z}]$ is the order-$z$ R\' {e}nyi divergence of  two probability distributions $\rho_1(X)$ and $\rho_2(X)$ \cite{Renyi1961,Erven2014}. From Equation \eqref{c1} to \eqref{c2} we have applied the Liouville theorem $dX=dX_0$ and Equation \eqref{crooks1}. Identifying the dissipated work $W_{\text{diss}}=W-\Delta F$ in Equation \eqref{c1}-\eqref{c4}, we consequently obtain Equation \eqref{central} for a classical system. Now we give several comments on Equation \eqref{central} for classical system: \\
(1). It relates a fundamental quantity in thermodynamics, the dissipated work, to a key concept in information theory, R\'{e}nyi divergences of two nonequilibrium phase space density distributions. For $z=1$, Equation \eqref{central} returns to the Jarzynski equality \cite{Jarzynski1997}. \\
(2). The fluctuation of the dissipated work is independent of
time $t$ because the densities on the right hand side of Equation \eqref{central} can be evaluated at any intermediate time. While the fact that the dissipated work is independent of $t$ can also be seen from \cite{Kawai2007,Parrondo2009}.  This time independence is a consequence of the Liouville equation in Hamilton dynamics. \\
(3). It is an exact relation between the generating function of the dissipated work in a driven process and the R\'{e}nyi divergences between the phase space density of the forward and its time reversed process at any intermediate time of the experiment. Differentiating both sides of Equation \eqref{central} with respect to $z$ $n$ times with $n=1,2,3,\cdots$ and then fixing $z=0$, we obtain the various moments of the dissipated work,
\begin{eqnarray}\label{moments}
\langle W_{\text{diss}}^n \rangle=T^n\int dX\rho_F(X;t)\Bigg(\ln\frac{\rho_F(X;t)}{\rho_R(\Theta X;\tau-t)}\Bigg)^n,
\end{eqnarray}
where $n=1,2,3,\cdots$ and $T$ is the temperature. In particular for $n=1$,  the mean of the dissipation is \cite{Kawai2007,Jarzynski2006,Jarzynski2009,Parrondo2009,Lindblad1974}
\begin{eqnarray}
\langle W_{\text{diss}}\rangle&=&T\int dX\rho_F(X;t)\ln\frac{\rho_F(X;t)}{\rho_R(\Theta X;\tau-t)},\\
&=&TD[\rho_F(X;t)||\rho_R(\Theta X;\tau-t)],
\end{eqnarray}
where $D[\rho_F(X;t)||\rho_R(\Theta X;\tau-t)]$ is the relative entropy \cite{Wehrl1978} between forward phase space density distributions and the reversed phase space density distributions. If a probability distribution $P(W)$ has finite moments of all orders $\langle W^n\rangle$ ($n$ from $0$ to $\infty$) and $\sum_{k=0}^{\infty}a_k\langle W^k\rangle/k!$ has any positive radius of convergence, then all the moments uniquely determine the distribution \cite{Bill1995}. In this case, the characteristic function of the distribution is given by,
\begin{eqnarray}
G(u)=\langle e^{iuW}\rangle=\sum_{n=0}^{\infty}\frac{(iu)^n}{n!}\langle W^n\rangle.
\end{eqnarray}
Whose Fourier transform gives the probability distribution. Thus Equation \eqref{moments} provides a means to obtain the probability distribution of dissipated work from non-equilibrium phase space density distributions. \\
(4). For some special values of $z$, the R\'{e}nyi divergence reduces to distance measures. For $z=1/2$, we have
\begin{eqnarray}
\langle e^{-\beta W_{\text{diss}}/2}\rangle&=&1-\frac{1}{2}D_H^2[\rho_R(\Theta X;\tau-t),\rho_F(X;t)].
\end{eqnarray}
where $D_H^2[P,Q]=\sum_{i=1}^n(\sqrt{p_i}-\sqrt{q_i})^2$ is the squared Hellinger distance \cite{Gibbs2002} of two distributions $P=\{p_1,p_2,\cdots,p_n\}$ and $Q=\{q_1,q_2,\cdots,q_n\}$.

\begin{figure}
\begin{center}
\includegraphics[scale=0.35]{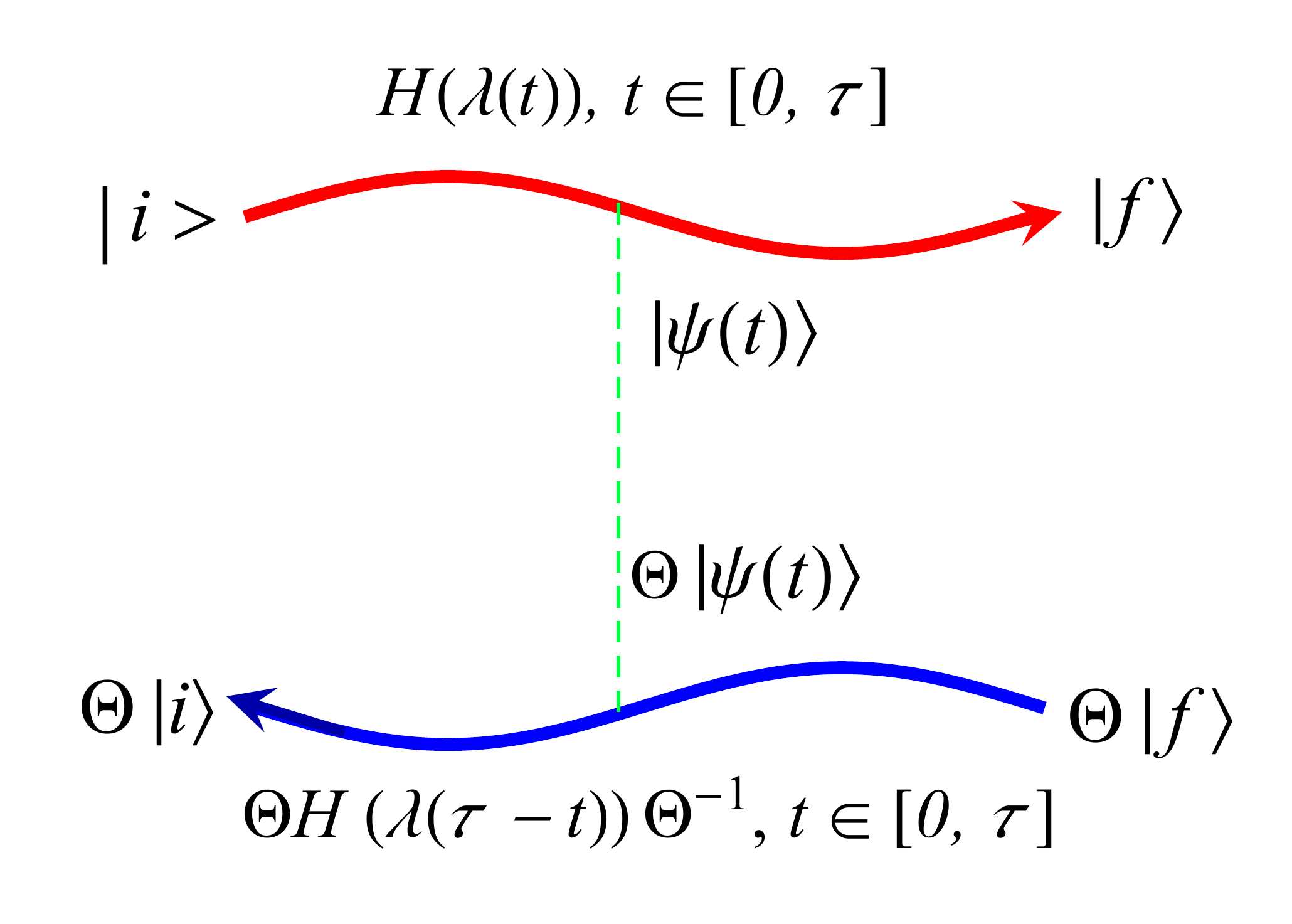}
\end{center}
\caption{(color online). Schematic illustration of the time reversal symmetry for quantum system with time-dependent Hamiltonian. The upper red line with arrow towards right denotes the forward process: it starts at an arbitrary initial state $|i\rangle$ and evolves in time under the unitary evolution generated by $H(\lambda(t')), t'\in[0,\tau]$ for time duration $\tau$. Then the state at time $t$ is $|\psi(t)\rangle=\mathcal{T}\exp[-i\int_0^tdt'H(\lambda(t'))]|i\rangle$ and finally becomes $|f\rangle$ at time $\tau$. The lower blue line with arrow towards left is the time reversed process: it starts at $\Theta |f\rangle$ (time reversed state of $|f\rangle$) and evolves under Hamiltonian $\Theta H(\lambda(\tau-t))\Theta^{-1}, t\in[0,\tau]$ for time duration $\tau$. Then the state at time $\tau-t$ is $\Theta|\psi(t)\rangle$ and finally becomes $\Theta |i\rangle$ at time $\tau$. }
\label{fig:epsart1}
\end{figure}

\emph{Non-equilibrium quantum thermodynamics}-
Let us consider a finite quantum system governed by a Hamiltonian $H(\lambda)$ and $\lambda$ is a parameter controlled by an external agent. We illustrate the time reversal symmetry for quantum system under time-dependent driving in Figure 2.

Let us first define the \emph{forward process in quantum system} under time-dependent driving. In the forward process, we initialize the quantum system in canonical equilibrium state at inverse temperature $\beta=1/T$ at a fixed value of control parameter $\lambda_i$, which is described by the density matrix $\rho_F(0)=e^{-\beta H(\lambda_i)}/Z(\beta,\lambda_i)$ with $Z(\beta,\lambda_i)=\text{Tr}[e^{-\beta H(\lambda_i)}]$ being the canonical partition function. Then we isolate the system and drive it by the Hamiltonian $H(\lambda(t))$ for a time duration $\tau$, where the force protocol $\lambda(t), t\in[0,\tau]$ brings the parameter from $\lambda_i$ at $t=0$ to $\lambda_f$ at a later time $\tau$. Then the state at $t$ in the forward process is given by
\begin{eqnarray}
\rho_F(t)&=&U_F(t,0)\rho_F(0)U_F^{\dagger}(t,0),
\end{eqnarray}
where $U_F(t,0)\equiv \mathcal{T}e^{-i\int_0^tdt'H(\lambda(t))}$ with $\mathcal{T}$ being the time ordering operator. Of course usually $\rho_F(t)\neq e^{-\beta H[\lambda(t)]}/Z[\beta,\lambda(t)]$.  Work in quantum system is defined by two projective measurements \cite{Talker2007,RMP2011}.
We assume, for any $\lambda$, $H(\lambda)|n_{\gamma}(\lambda)\rangle=E_n(\lambda)|n_{\gamma}(\lambda)\rangle$ and the symbol $n$ labels eigenenergy and $\gamma$ denotes further quantum numbers to specify an energy eigenstate in case of $g_n$-fold degeneracy. At $t=0$, the first projective measurement of $H(\lambda_i)$ is performed with outcome $E_n(\lambda_i)$ with probability $p_n(0)=g_ne^{-\beta E_n(\lambda_i)}/Z(\beta,\lambda_i)$. Simultaneously the initial equilibrium state projects into the state, $\sigma_n=\Pi_n(\lambda_i)\rho_F(0)\Pi_n(\lambda_i)/p_n(0)$ with $\Pi_n(\lambda)\equiv\sum_{\gamma}|n_{\gamma}(\lambda)\rangle\langle n_{\gamma}(\lambda)|$. At $0<t<\tau$, the system is isolated and driven by a unitary evolution operator $U_F(\tau,0)=\mathcal{T}e^{-i\int_0^{\tau}H(\lambda(t))dt}$ and the state at $\tau$ is $\sigma_n(\tau)=U_F(\tau,0)\sigma_nU_F^{\dagger}(\tau,0)$. At $t=\tau$, the second projective measurement of $H(\lambda_f)$ yielding the eigenvalue $E_m(\lambda_f)$ with conditional probability $p_{m|n}(\tau)=\text{Tr}[\Pi_m(\lambda_f)\sigma_n(\tau)]$ is performed. Work is defined by difference of energy measurements. So the probability of obtaining $E_n(\lambda_i)$ for the first measurement and followed by obtaining $E_m(\lambda_f)$ in the second measurement is $p_n(0)p_{m|n}(\tau)$. Thus the work distribution in the forward driven process is given by \cite{Talker2007,RMP2011}
\begin{eqnarray}\label{QWD}
P_F(W)=\sum_{m,n}p_n(0)p_{m|n}(\tau)\delta[W-E_m(\lambda_f)+E_n(\lambda_i)].
\end{eqnarray}
The quantum work distribution $P_F(W)$ encodes the fluctuations in the work
that arise from thermal statistics and from quantum
measurement statistics over many identical realizations
of the protocol.

Now we define the \emph{reversed process in quantum system} under time-dependent driving. In the reversed process, we initialize the quantum system in the time reversed state of the canonical equilibrium state at inverse temperature $\beta=1/T$ at value $\lambda_f$ of the control parameter, $\rho_R(0)=\Theta e^{-\beta H(\lambda_f)}\Theta^{-1}/Z(\beta,\lambda_f)$ with $Z(\beta,\lambda_f)=\text{Tr}[e^{-\beta H(\lambda_f)}]$ being the canonical partition function. Then we drive the system by the Hamiltonian in the reversed process $H_R(t)=\Theta H(\lambda(\tau-t))\Theta^{-1}$ for a time duration $\tau$ which brings the force parameter from $\lambda_f$ at $t=0$ to $\lambda_i$ at a later time $\tau$. The time evolution operator for the forward driven process and its the reversed process are related by \cite{Andri2008}
\begin{eqnarray}\label{frrelation}
U_R(t,0)=\Theta U_F^{\dagger}(\tau,\tau-t)\Theta^{-1}.
\end{eqnarray}
Then the state at $t$ in the reversed process is given by
\begin{eqnarray}\label{reversematrix}
\rho_R(t)&=&U_R(t,0)\rho_R(0)U_R^{\dagger}(t,0),
\end{eqnarray}
where $U_R(t,0)\equiv \mathcal{T}e^{-i\int_0^tdt'H_R(t)}$. Although $\rho_F(t)$ and $\rho_R(t)$ are far from equilibrium states, they satisfy the following lemma due to time reversal symmetry in the forward process and the reversed process:

\emph{Lemma}-The density matrices in the forward driven process at arbitrary time $t\in[0,\tau]$ and its time reversed processes at time $\tau-t$ satisfy, for any finite real numbers $a,b\in \Re,$
\begin{eqnarray}\label{lemma}
\text{Tr}\Big[\Big(\Theta^{-1}\rho_R(\tau-t)\Theta\Big)^a\Big(\rho_F(t)\Big)^b\Big]&=&
\text{Tr}\Big[\Big(\Theta^{-1}\rho_R(\tau)\Theta\Big)^a\Big(\rho_F(0)\Big)^b\Big].\nonumber\\
\end{eqnarray}
\\
\emph{Proof}: From Equation \eqref{frrelation} and \eqref{reversematrix}, we have
\begin{eqnarray}
\Theta^{-1}\rho_R(\tau-t)\Theta&=&\Theta^{-1}U_R(\tau-t,0)\rho_R(0)U_R^{\dagger}(\tau-t,0)\Theta,\\
&=&U_F^{\dagger}(\tau,t)\Theta^{-1}\rho_R(0)\Theta U_F(\tau,t).
\end{eqnarray}
Then \begin{eqnarray}
&&(\Theta^{-1}\rho_R(\tau-t)\Theta)^a(\rho_F(t))^b,\\
&=&U_F^{\dagger}(\tau,t)\Big(\Theta^{-1}\rho_R(0)\Theta\Big)^a U_F(\tau,t)U_F(t,0)\rho_F^b(0)U_F^{\dagger}(t,0),\\
&=&U_F(t,0)U_F^{\dagger}(\tau,0)(\Theta^{-1}\rho_R(0)\Theta)^a U_F(\tau,0)\rho_F^b(0)U_F^{\dagger}(t,0),\\
&=&U_F(t,0)(\Theta^{-1}\rho_R(\tau)\Theta)^a(\rho_F(0))^bU_F^{\dagger}(t,0),
\end{eqnarray}
which means $(\Theta^{-1}\rho_R(\tau-t)\Theta)^a(\rho_F(t))^b$ and $(\Theta^{-1}\rho_R(\tau)\Theta)^a(\rho_F(0))^b$ are related to each other by a unitary transformation $U_F(t,0)$. They must be equal under the trace. Thus we have proved Equation \eqref{lemma}.

From the definition of quantum work distribution, Equation \eqref{QWD} and the Lemma proved above, Equation \eqref{lemma}, we have
\begin{eqnarray}
&&\langle \Big(e^{-\beta W}\Big)^z\rangle_F\label{q1}\\
&=&Z_i^{-1}\text{Tr}\Big[U_F(\tau,0)e^{-\beta(1-z)H_i}U_F^{\dagger}(\tau,0)e^{-\beta zH_f}\Big],\\
&=&\frac{Z_f^z}{Z_i^z}\text{Tr}\bigg[U_F(\tau,0)\Big(\rho_F(0)\Big)^{1-z}U_F^{\dagger}(\tau,0)\Big(\Theta^{-1}\rho_R(0)\Theta\Big)^{z}\bigg],\\
&=&\frac{Z_f^z}{Z_i^z}\text{Tr}\bigg[\Big(\rho_F(\tau)\Big)^{1-z}\Big(\Theta^{-1}\rho_R(0)\Theta\Big)^{z}\bigg],\label{d1}\\
&=&\frac{Z_f^z}{Z_i^z}\text{Tr}\bigg[\Big(\rho_F(t)\Big)^{1-z}\Big(\Theta^{-1}\rho_R(\tau-t)\Theta\Big)^{z}\bigg],\label{d2}\\
&=& e^{-\beta z\Delta F}e^{(z-1)S_z[\Theta^{-1}\rho_R(\tau-t)\Theta||\rho_F(t)]}\label{Qfree}.
\end{eqnarray}
Here $z$ is a finite real number and $\Delta F\equiv F[\beta,\lambda_f]-F[\beta,\lambda_i]$. From Equation \eqref{d1} to Equation \eqref{d2}, we have used the lemma proved above. In the last step, we have made use of definition of the order-$z$ quantum R\'{e}nyi divergence of two density matrices $\rho_1$ and $\rho_2$ \cite{Beigi2013,Lennert2013}, $S_z(\rho_1||\rho_2)\equiv\frac{1}{z-1}\ln[\text{Tr}[\rho_1^{z}\rho_2^{1-z}]]$, which is information theoretic generalization of standard relative entropy \cite{Renyi1961}. If we identify $W-\Delta F$ as the dissipated work $W_{\text{diss}}$ in Equation \eqref{q1} and \eqref{Qfree}, we therefore obtain Equation \eqref{central} for quantum system. Now we make several comments on Equation \eqref{central} for quantum system:\\
(1). It relates a fundamental quantity in quantum thermodynamics, the dissipated work, to a fundamental concept in quantum information theory, the quantum R\'{e}nyi divergences between the nonequilibrium density matrix in the forward process at arbitrary time $t$ and the density matrix in the reversed process at any time $\tau-t$. \\
(2). The fluctuation of the dissipated work in quantum system is independent of time $t$ because the density matrices on the right hand side of Equation \eqref{central} can be evaluated at any intermediate time. While the fact that it is independent of $t$ can also be seen from \cite{Parrondo2009,Deffner2010}.  This time independence is a consequence of the time reversal symmetry in driven process.\\
(3). It is an exact relation between the generating function of the dissipated work and R\'{e}nyi divergences between a non-equilibrium density matrix in the forward process at any intermediate time $t$ and the density matrix in the reversed process at time $\tau-t$. Differentiating both sides of Equation \eqref{central} with respect to $z$ $n$ times with $n=1,2,3,\cdots$ and then setting $z=0$, we obtain the various moments of the dissipated work for quantum system under time-dependent driving,
\begin{eqnarray}
\langle W_{\text{diss}}^n\rangle_F&=&T^n\text{Tr}\Big[\rho_F(t)\mathcal{T}_n\Big(\ln[\rho_F(t)]-\ln[\Theta^{-1}\rho_R(\tau-t)\Theta]\Big)^n\Big],\nonumber\\\label{momentsquan}
\end{eqnarray}
where $T$ is the temperature, $n=1,2,3,\cdots$ and $\mathcal{T}_n$ is an ordering operator which sorts that in each term of the binomial expansion of $\Big(\ln[\rho_F(t)]-\ln[\Theta^{-1}\rho_R(\tau-t)\Theta]\Big)^n$, $\ln[\rho_F(t)]$ always sits on the left of $\ln[\Theta^{-1}\rho_R(\tau-t)\Theta]$.
In particular for $n=1$, it is \cite{Parrondo2009,Deffner2010,Vedral2012}
\begin{eqnarray}
\langle W_{\text{diss}}\rangle&=&T\text{Tr}\Big[\rho_F(t)\Big(\ln[\rho_F(t)]-\ln[\Theta^{-1}\rho_R(\tau-t)\Theta]\Big)\Big],\nonumber\\
&=&TD[\rho_F(t)||\Theta^{-1}\rho_R(\tau-t)\Theta],\label{qdiss}
\end{eqnarray}
where $D[\rho_F(X;t)||\rho_R(\Theta X;\tau-t)]$ is the von Neumann relative entropy \cite{Umegaki1962,Wehrl1978} between density matrix in the forward process at arbitrary time $t\in[0,\tau]$ and the density matrix in the reversed process at time $\tau-t$. Recently this result was experimentally demonstrated by using a nuclear magnetic resonance set-up that allows for measuring the non-equilibrium entropy produced in an isolated spin-1/2 system following fast quenches of an external magnetic field \cite{Serra2015}. As in the classical case, if a probability distribution $P(W)$ has finite moments of all orders $\langle W^n\rangle$ ($n$ from $0$ to $\infty$) and $\sum_{k=0}^{\infty}a_k\langle W^k\rangle/k!$ has any positive radius of convergence, then all the moments uniquely determine the distribution \cite{Bill1995}. Thus Equation \eqref{momentsquan} provides a means to obtain distribution of dissipated work for quantum system driven arbitrarily far from equilibrium from the non-equilibrium density matrices.\\
(4). For $z=1/2$, the order-1/2 R\'{e}nyi divergence is related to the squared Hellinger distance \cite{Gibbs2002}, $D_H^2[\rho_1,\rho_2]=\text{Tr}[(\sqrt{\rho_1}-\sqrt{\rho_2})^2]$ for two density matrices $\rho_1$ and $\rho_2$. We thus have
\begin{eqnarray}
\langle e^{-\beta W_{\text{diss}}/2}\rangle&=&1-\frac{1}{2}D_H^2[\rho_R(\Theta X;\tau-t),\rho_F(X;t)].
\end{eqnarray}

\label{conclusion} In summary, we have established an exact relation which connects a fundamental quantity in non-equilibrium thermodynamics, the dissipated work in a system driven arbitrarily far from equilibrium, to a fundamental concept in information theory, R\'{e}nyi divergences. We find that the generating function of the dissipated work under an arbitrary time-dependent driving is related to the R\'{e}nyi-divergences between a non-equilibrium state in the driven process at an arbitrary intermediate time and a non-equilibrium state in its time reversed process at an arbitrary intermediate time. This relation is universally applicable to both finite classical system and finite quantum system, arbitrarily far from equilibrium. In this work, we studied the case that the system is isolated from the bath in the time-dependent driving process, it would be interesting to study whether the results still hold if the system and bath are coupled in the course of driving.

\begin{acknowledgements}
This work was supported by an Alexander von Humboldt Professorship and the EU Projects EQUAM and SIQS.
\end{acknowledgements}

\end{document}